\documentclass{article}

\pdfoutput=1

\usepackage[pdftex]{graphicx}
\usepackage{subfigure}
\usepackage{amsmath, amsthm, amssymb}
\usepackage{amsfonts}
\usepackage{color}
\usepackage{bm}
\usepackage{bbm}
\usepackage{tikz}
\usepackage{natbib}
\usepackage{caption}
\usepackage{algorithm}
\usepackage{algorithmic}

\theoremstyle{plain}

\newcommand{\phimax}{\ensuremath{\varphi_0}}

\newcommand{\domain}[4]
{
\begin{tikzpicture}[xscale=1.25,yscale=1.25,
    declare function={
      threshold=0.01;
      term1=#1-#2*#3;
      term2=#2-#1*#3;
      term3=#3-#2*#1;
      temp1=1-#1*#1;
      temp2=1-#2*#2;
      temp3=1-#3*#3;
      sqr=sqrt(temp3);
      phimax=acos(-#3)*pi/180;
      limit=-term2/sqrt(temp1*temp3);
      thetamax=acos(limit)*pi/180;
      denom(\x)=ifthenelse(\x<.99*phimax,sqr*cos(\x r)+#3*sin(\x r),threshold);
      scale(\x) = sin(\x r)/denom(\x);
funct(\x)=-(term1+scale(\x)*term2)/    
sqrt(temp3*(temp2-2*scale(\x)*term3+scale(\x)*scale(\x)*temp1))         ;
      bound(\x)
= acos(ifthenelse(denom(\x)>threshold,funct(\x), limit))*pi/180;
    }
    ]

    \pgfmathsetmacro{\phimax}{phimax};
    \pgfmathsetmacro{\thetamax}{thetamax}
    \pgfmathsetmacro{\thetamin}{bound(0)}

        \draw[help lines] (0,0) grid (pi,pi); 
    \draw [<->] (pi,0) node[below]{$\varphi$} -- (0,0) -- (0,pi) node[left]{$\theta$};
    
    \draw [->,ultra thick] (0,0) -- ({\phimax/2},0); 
    \draw [ultra thick] ({\phimax/2},0) -- ({\phimax},0);
    \node [below] at ({\phimax},-0.1) {$\varphi_0$}; 
    \node [below] at ({\phimax/2},-0.1) {$C_1$}; 
  
        \draw [->,ultra thick] (\phimax,0) -- (\phimax,{\thetamax/2}); 
    \draw [ultra thick] (\phimax,{\thetamax/2}) -- (\phimax,\thetamax); 
    \node [right] at (\phimax+0.1,{\thetamax/2}) {$C_2$}; 
    
        \draw [->,ultra thick] (0,\thetamin) -- (0,\thetamin/2); 
    \draw [ultra thick] (0,\thetamin/2) -- (0,0); 
    \node [left] at (-0.1,{\thetamin/2}) {$C_4$};

        \draw [->,ultra thick,domain={\phimax}:{\phimax/2}] plot(\x,{bound(\x)});
    \draw [ultra thick,domain={\phimax/2}:0] plot(\x,{bound(\x)});
    \node [above] at ({\phimax/2},{bound(\phimax/2)}) {$C_3$};
    \node at ({\phimax/2},{(\thetamin+\thetamax)/4}) {$\Omega$};
\end{tikzpicture}
\begin{center}
\captionof{figure}{$\rho_{xy}=#3,\quad \rho_{xz}=#2,\quad \rho_{yz}=#1$}
\label{fig:#4}
\end{center}
}
\begin{document}

\title{A structural approach to pricing credit default swaps with credit and
debt value adjustments}
\author{Alexander Lipton, Ioana Savescu \\
Bank of America Merrill Lynch \\
Imperial College, London, UK}
\maketitle

\begin{abstract}
A multi-dimensional extension of the structural default model with firms'
values driven by diffusion processes with Marshall-Olkin-inspired
correlation structure is presented. Semi-analytical methods for solving the
forward calibration problem and backward pricing problem in three dimensions
are developed. The model is used to analyze bilateral counterparty risk
for credit default swaps and evaluate the corresponding credit and debt
value adjustments.
\end{abstract}

\section{Introduction}

The recent financial crisis has profoundly changed the nature of credit markets
in general and correlation trading in particular. The focus has shifted from
complicated products, such as bespoke collateralized debt obligations
(CDOs), CDOs-Squared, etc., towards simpler products, such as credit
indices, collateralized credit default swaps (CDSs), funded single name
credit-linked notes (CLNs), CDSs collateralized by a risky bond, etc, for which risks are somewhat easier to understand and model. However,
as recent events have shown, if not properly managed, the trading of even these
relatively simple products can cause big losses. More details can be found in \citet{lipton2011oxford}.

During the crisis, it has become apparent that proper accounting for
counterparty risk in the valuation of over-the-counter (OTC) derivatives is
extremely important, especially in view of the fact that some protection
sellers, such as mono-line insurers and investment banks, have experienced
sharply elevated default probabilities or even default events, the case of
Lehman Brothers being the prime example. Counterparty credit risk is the risk
that a party to a financial contract will default prior to its expiration
and will not fulfill all of its obligations. In principle, only OTC
contracts privately negotiated between counterparties are subject to
counterparty risk.

The structural model first introduced by Merton is one of the two standard
models used for pricing single-name CDSs, the other one being the
reduced-form model. Extensions of the structural model to the
two-dimensional case have been proposed by \citet{Zhou2001}, %
\citet{Patras2006}, \citet{Valuzis2008}, among others, who considered
correlated log-normal dynamics for two firms and derived analytical formulas
for their joint survival probability using the eigenvalue expansion
technique; see also \citet{Lipton2001}, \citet{he1998double}, where an
identical technique was used in a different context. Two-dimensional
structural models have been successfully used for the estimation of the
credit value adjustment (CVA), and the debt value adjustment (DVA) for CDSs
(see, e.g., \citet{LiptonSepp2009}, \citet{ScailletPatras2008}).

In order to compute the CVA (DVA), one needs to study the joint evolution of
the assets of the reference name and the protection seller (buyer), provided
that the corresponding CDS is viewed from the standpoint of the protection
buyer. Clearly, in order to calculate the CVA and DVA for a CDS \textit{%
simultaneously and consistently}, one needs to consider three-dimensional
structural models and study the joint evolution of the assets of the
reference name, the protection seller and the protection buyer. This paper
extends the results of \citet{LiptonSepp2009} by considering correlated
log-normal dynamics for three firms and computing their joint survival
probability. The corresponding problem is solved by using the eigenfunction
expansion technique combined with the finite element method to obtain a
semi-analytical expression for the Green's function. Once the Green's
function is known, both CVA and DVA corrections for a CDS can be computed in
a consistent manner. The power of the proposed technique is illustrated by
considering some realistic examples of pricing CDSs sold by risky sellers to
risky buyers. As might be expected, counterparty credit effects have great
impact on the value of a CDS contract.

\section{CVA/DVA for CDSs}

\label{sect:CVA}

A CDS is a contract in which the protection buyer (PB) agrees to pay a periodic
coupon $c$ to a protection seller (PS) in exchange for a potential cashflow
in the event of a default of the reference name (RN) of the swap before the
maturity of the contract $T$. The value can be naturally decomposed
into a coupon leg (CL) and a default leg (DL). We denote by $\tau^{RN}$ the default
time of the reference name and by $R_{RN}$ its recovery, and we have (from
the protection buyer's point of view): 
\begin{align*}
CL_{t}& =-\mathbb{E}\left[ \left. \textstyle{\sum_{T_{i}}{cD\left(
t,T_{i}\right) \mathbbm{1}_{\left\{ T_{i}\leq \tau ^{RN}\right\} }\Delta T}}%
\right\vert \mathcal{F}_{t}\right] , \\
DL_{t}& =\mathbb{E}\left[ \left. \left( 1-R_{RN}\right) D(t,\tau^{RN})%
\mathbbm{1}_{\left\{ t<\tau^{RN}<T\right\} }\right\vert \mathcal{F}_{t}%
\right] ,
\end{align*}%
where $T_{i}$ are the coupon payment dates and $D(t,T)$ is the price of a
zero-coupon bond with maturity $T$.

In order to simplify the formulas we denote by $CF(t,T)$ the sum of all discounted
contractual cashflows between $t$ and the maturity $T$ (both coupon leg and
default leg), and write the value $V_t$ of the CDS as: $V_t = 
\mathbb{E}\left[ \left. CF(t,T) \right| \mathcal{F}_{t} \right] $.

We suppose now that the protection seller can default but consider the
protection buyer risk free, and denote by $\tilde{V}_{t}$ the value of the
derivative in this case: 
\begin{align*}
\tilde{V}_{t}=& \mathbb{E}\left[ \left. CF(t,T)\mathbbm{1}_{\left\{ \tau
^{PS}>\min \{T,\tau ^{RN}\}\right\} }\right\vert \mathcal{F}_{t}\right] \\
& +\mathbb{E}\left[ \left. \mathbbm{1}_{\left\{ \tau ^{PS}<\min \{T,\tau
^{RN}\}\right\} }\left[ CF(t,\tau ^{\scriptscriptstyle{PS}})+D(t,\tau ^{%
\scriptscriptstyle{PS}})\left( R_{PS}V_{\tau ^{PS}}^{+}+V_{\tau
^{PS}}^{-}\right) \right] \right\vert \mathcal{F}_{t}\right] ,
\end{align*}%
where $\tau ^{PS}$ denotes the default time of the protection seller; as usual, $V^{\pm} = \pm \max\left(0, \pm V\right)$. The
term Credit Value Adjustment (CVA) will refer to the additional cost to
account for the possibility of the counterparty's default and is defined as $%
\text{CVA}=V_{t}-\tilde{V}_{t}$: 
\begin{equation}
CVA=\left( 1-R_{PS}\right) \mathbb{E}\left[ \left. \mathbbm{1}_{\left\{ \tau
^{PS}<\min \{T,\tau ^{RN}\}\right\} }D(t,\tau ^{PS})V_{\tau
^{PS}}^{+}\right\vert \mathcal{F}_{t}\right] .  \label{eq:CVA_bi}
\end{equation}%
Similarly we can consider the case where the protection buyer is risky but
the protection seller is risk free. The term Debt Valuation Adjustment (DVA)
represents the additional cost to account for one's own default ($\tau
^{PB}$ denotes the default time of the protection buyer): 
\begin{equation}
DVA=\left( 1-R_{PB}\right) \mathbb{E}\left[ \left. \mathbbm{1}_{\left\{ \tau
^{PB}<\min \{T,\tau ^{RN}\}\right\} }D(t,\tau ^{PB})V_{\tau
^{PB}}^{-}\right\vert \mathcal{F}_{t}\right] .  \label{eq:DVA_bi}
\end{equation}

Given recent events, we can no longer suppose that one of the counterparties
is risk free. The Basel II documentation makes reference to a bilateral
counterparty risk, in which the default of both counterparties in
the derivative contract are subject to default risk. One of the advantages
of considering the bilateral CVA is the symmetry it introduces in pricing:
the two counterparties will now agree on the price of the derivative (for a
detailed discussion on this see for example \cite{BrigoCapponi2008}). If $\tau $ denotes the
minimum of the two default times: $\tau =\text{min}\{\tau^{PS},\tau^{PB}\}$, then 
\begin{align*}
\tilde{V}_{t}=& \mathbb{E}\left[ CF(t,T)\mathbbm{1}_{\left\{ \tau >T\right\}
}+\right. \\
& +\mathbbm{1}_{\left\{ \tau =\tau ^{PS}<T\right\} }\left( CF(t,\tau
^{PS})+D(t,\tau ^{PS})R_{PS}V_{\tau ^{PS}}^{+}+D(t,\tau ^{PS})V_{\tau
^{PS}}^{-}\right) + \\
& +\left. \mathbbm{1}_{\left\{ \tau =\tau ^{PB}<T\right\} }\left( CF(t,\tau
^{PB})+D(t,\tau ^{PB})R_{PB}V_{\tau ^{PB}}^{-}+D(t,\tau ^{PB})V_{\tau
^{PB}}^{+}\right) \right] .
\end{align*}

In the case where both counterparties are considered risky, bilateral CVA is
the combination of the two adjustments (CVA and DVA): 
\begin{equation}
CVA=\left( 1-R_{PS}\right) \mathbb{E}\left[ \left. \mathbbm{1}_{\left\{ \tau
^{PS}<min\{T,\tau ^{PB},\tau ^{RN}\}\right\} }D(t,\tau ^{PS})V_{\tau
^{PS}}^{+}\right\vert \mathcal{F}_{t}\right] ,  \label{eq:CVA_tri}
\end{equation}%
\begin{equation}
DVA=\left( 1-R_{PB}\right) \mathbb{E}\left[ \left. \mathbbm{1}_{\left\{ \tau
^{PB}<min\{T,\tau ^{PS},\tau ^{RN}\}\right\} }D(t,\tau ^{PB})V_{\tau
^{PB}}^{-}\right\vert \mathcal{F}_{t}\right] .  \label{eq:DVA_tri}
\end{equation}%
We emphasize that expressions (\ref{eq:CVA_bi}), (\ref{eq:DVA_bi}) and (\ref%
{eq:CVA_tri}), (\ref{eq:DVA_tri}) are not identical.

\section{Structural model framework}

\label{sect:framework}

We assume that the default and counterparty risk can be hedged, so that we can
work with the risk neutral pricing measure denoted by $\mathbb{Q}$. We also
assume a risk-free deterministic rate of return $\varrho _{t}$. We start
with the firm's asset value dynamics, which we denote by $a_{t},$ and assume (similar to the setup in \citet{LiptonSepp2009}) that it is driven by the following jump-diffusion dynamics under $\mathbb{Q}$: 
\begin{equation}
\frac{da_{t}}{a_{t}}=\left( \varrho _{t}-\zeta _{t}-\lambda _{t}\kappa
\right) dt+\sigma _{t}dW_{t}+\left( e^{j}-1\right) dN_{t},
\label{eq:assetDyn}
\end{equation}%
where $\zeta _{t}$ is the dividend rate, $W_{t}$ is a standard Brownian
motion, $\sigma _{t}$ is the deterministic volatility, $N_{t}$ is a Poisson
process independent of $W_{t}$, $\lambda _{t}$ its intensity, $j$ is the
jump amplitude and $\kappa $ is the jump compensator. We assume that the
firm defaults when its asset value becomes less than a fraction of its
debt per share and that the default barrier of the firm is a deterministic
function of time given by $l_{t}=l_{0}E_{t}$, where 
\begin{equation*}
\textstyle{E_{t}=\exp \left( \int_{0}^{t}{\left( \varrho _{u}-\zeta
_{u}-\lambda _{u}\kappa -\frac{1}{2}\sigma _{u}^{2}\right) du}\right) ,}
\end{equation*}%
and $l_{0}=RL_{0}$. Here $R$ is the recovery on the firm's liabilities and $%
L_{0}$ is its total debt per share. We consider the firm's equity price per
share $s_{t}$ and we assume that it is given by $s_{t}=a_{t}-l_{t}$, for $%
t<\tau $ and $0$ for $t\geq \tau $ where $\tau $ is the default time. The
solution of the stochastic differential equation \eqref{eq:assetDyn} can be
written as a product of a deterministic part and a stochastic exponent $%
a_{t}=l_{0}E_{t}e^{\sigma _{t}x_{t}}$, where the stochastic factor $x_{t}$
has the following dynamics under $\mathbb{Q}$: 
\begin{equation}
dx_{t}=dW_{t}+\frac{j}{\sigma _{t}}dN_{t}, \quad x_0 > 0, \label{eq:x_t}
\end{equation}%
with $x_{t}$ representing the \textquotedblleft relative
distance\textquotedblright\ of the asset value from the default barrier. The
default event occurs at the first time $\tau $ when $x_{\tau }$ becomes
negative, so that the default barrier is fixed at zero.

Introducing jumps in the dynamics of the asset value allows us to calibrate
to CDS market spreads even for short maturities. In a framework without
jumps it is well known that a good calibration of the short end of the curve
is impossible. However, this simpler framework allows for analytical solutions in some
cases which provide insight into the problem as well as
a good benchmark for the more general case with jumps. In this paper
we therefore focus on the simplified case without jumps.

For the multi-dimensional case we consider that the process for the relative distance to
default for each of the entities of interest has a similar dynamics
to equation \eqref{eq:x_t} but also in the simplified framework without the jump
component, and we correlate the diffusions in the usual way by assuming $%
d\langle W_t^i, W_t^j \rangle = \rho_{ij}dt$.

\section{Two dimensional case}

\label{sect:2D}

The one dimensional case of the standard CDS where both
counterparties are non-risky admits well-known analytical solutions. We therefore turn our focus directly to the two
dimensional problem where we need to model the dynamics of the asset values
of the reference name and protection seller simultaneously, while
considering the protection buyer to be non-risky. We work with the processes 
$x_{t}$ and $y_{t}$ for the relative distance from the default barrier in time for
each of the two entities considered. These processes have the following
dynamics $dx_{t}=dW_{t}^{x},\ \ dy_{t}=dW_{t}^{y}$, where the Brownian
motions are correlated with correlation $\rho _{xy}$.

\subsection{Pricing problem}

\label{sect:2D_PrEq}

The general pricing equation in this framework is given by: 
\begin{equation}
V_{t}+\frac{1}{2}V_{xx}+\frac{1}{2}V_{yy}+\rho _{xy}V_{xy}-\varrho V=0.
\label{eq:to_solve}
\end{equation}%
We consider the following function $U(t,x,y)=e^{\varrho (T-t)}V(t,x,y)$ and
make a change of variable that allows us to eliminate the cross derivatives: 
\begin{equation}
\alpha (x,y)=x,\quad \quad \beta (x,y)=-\frac{1}{\bar{\rho}_{xy}}\left( \rho
_{xy}x-y\right) ,  \label{eq:2D_1stVars}
\end{equation}%
where we have used the usual notation $\bar{\rho}_{xy}=\sqrt{1-\rho
_{xy}^{2}}$. The domain in which the equation has to be solved has changed from
the positive quadrant to the interior of an angle. This angle is
characterized by $\cos (\varphi _{0})=-\rho _{xy}$, so if $\rho _{xy}>0$,
the angle is blunt. In order to take advantage of the symmetry of the
domain, we make a second change of variables and pass to polar coordinates, $%
\left( \alpha ,\beta \right) =\left( -r\sin \left( \varphi -\varphi
_{0}\right) ,r\cos \left( \varphi -\varphi _{0}\right) \right) $.

\subsection{Green's function}

The Green's function solves the forward problem (where $\tau =T-t$): 
\begin{eqnarray}
&&G_{\tau }-\frac{1}{2}\left( G_{rr}+\frac{1}{r}G_{r}+\frac{1}{r^{2}}%
G_{\varphi \varphi }\right) =0,  \label{eq:2D_Green} \\
&&G(0,r,\varphi )=\frac{1}{r^{\prime }}\delta (r-r^{\prime })\delta (\varphi
-\varphi ^{\prime })  \notag \\
&&G\left( \tau ,r,0\right) =0\quad G(\tau ,r,\varphi _{0})=0\quad G(\tau
,0,\varphi )=0\quad G(\tau ,r,\varphi )\xrightarrow[r\to \infty]{}0.  \notag
\end{eqnarray}

Two possible methods can be applied in order to obtain the solution for the
Green's function: the eigenvalue expansion method and the method of images.
The solution using the first method is well known and was first introduced in \citet{Zhou2001}, \citet{Lipton2001}, %
\citet{he1998double}. The resulting formula for the Green's function is: 
\begin{equation}
G(\tau ,r,\varphi |0,r^{\prime },\varphi ^{\prime })=\frac{2e^{-\frac{%
r^{2}+r^{\prime 2}}{2\tau }}}{\varphi _{0}\tau }\sum_{n=1}^{\infty }{I_{%
\frac{n\pi }{\varphi _{0}}}\left( \frac{rr^{\prime }}{\tau }\right) \sin {%
\left( \frac{n\pi \varphi }{\varphi _{0}}\right) }\sin {\left( \frac{n\pi
\varphi ^{\prime }}{\varphi _{0}}\right) .}}  \label{eq:Green2D}
\end{equation}

A solution through the method of images was announced by Lipton in 2008
at a SIAM meeting, and briefly discussed in \citet{LiptonSepp2009}. Here we
present an improved version. We first need to find the solution to equation %
\eqref{eq:2D_Green} with the same initial condition but with non-periodic
boundary conditions: 
\begin{equation*}
G(\tau ,0,\varphi )=0\quad \quad G(\tau ,r,\varphi )\xrightarrow[r\to
\infty]{}0\quad G\left( \tau ,r,\varphi \right) \xrightarrow[|\varphi| \to
\infty]{}0.
\end{equation*}

Using a Fourier transform technique, and in order to write the final expression in a more compact
form we introduce the following function $f\left(
p,q\right) $ where $p\geq 0,-\infty <q<\infty $: 
\begin{equation*}
f\left( p,q\right) =1-\frac{1}{2\pi }\int_{-\infty }^{\infty }\frac{%
e^{-p\left( \cosh \left( 2q\zeta \right) -\cos (q)\right) }}{\zeta ^{2}+%
\frac{1}{4}}d\zeta ,
\end{equation*}%
and its extension $h\left( p,q\right) $ defined as follows: 
\begin{equation*}
h\left( p,q\right) =\frac{1}{2}\left[ s_{+}f\left( p,\pi +q\right)
+s_{-}f\left( p,\pi -q\right) \right] ,
\end{equation*}%
where $s_{\pm }=\text{sign}\left( \pi \pm q \right) $. Then we can represent $G\left( t,r,\varphi |0,r^{\prime
},\varphi ^{\prime }\right) $ in the following form (which can be viewed as
a direct generalization of the one dimensional case): 
\begin{equation*}
G\left( \tau ,r,\varphi |0,r^{\prime },\varphi ^{\prime }\right) =\frac{1}{%
2\pi \tau }e^{-\frac{r^{2}+r^{\prime 2}-2\cos \left( \varphi -\varphi
^{\prime }\right) rr^{\prime }}{2\tau }}h\left( \frac{rr^{\prime }}{\tau }%
,\varphi -\varphi ^{\prime }\right) .
\end{equation*}

We can now apply the method of images and represent the fundamental solution
in the form 
\begin{equation}
\textstyle{G_{\varphi _{0}}\left( \tau ,r,\varphi |0,r^{\prime },\varphi
^{\prime }\right) =\sum\limits_{n=-\infty }^{\infty }\!\!\!\left[ G\left(
\tau ,r,\varphi |0,r^{\prime },\varphi ^{\prime }+2n\varphi _{0}\right)
-G\left( \tau ,r,\varphi |0,r^{\prime },-\varphi ^{\prime }+2n\varphi
_{0}\right) \right] .}  \label{eq:2D_ImagesG}
\end{equation}

The representation given in equation \eqref{eq:2D_ImagesG} gives, as expected,
exactly the same results for the Green's function as those obtained
through the eigenvalue expansion method. 

\subsection{Joint survival probability}

We denote by $Q(t,x,y)$ the joint survival probability of issuers $x$ and $y$
to a fixed maturity $T$. This solves the following pricing equation 
\begin{equation*}
Q_{t}+\frac{1}{2}Q_{xx}+\frac{1}{2}Q_{yy}+\rho _{xy}Q_{xy}=0,
\end{equation*}%
with final condition $Q(T,x,y)=1$ and boundary conditions $Q(t,x,0)=0$ and $%
Q(t,0,y)=0$. We use the expression for the Green's function obtained through
the eigenvalue expansion method and we obtain the following expression for
the survival probability in the new variables: 
\begin{equation*}
Q(t,r^{\prime },\varphi ^{\prime })=\frac{4}{\pi }\int_{0}^{\infty }\frac{{%
e^{-\frac{r^{2}+r^{\prime 2}}{2\tau }}}}{\tau }{\sum_{k=0}^{\infty }{\frac{1%
}{2k+1}I_{\frac{(2k+1)\pi }{\varphi _{0}}}\left( \frac{rr^{\prime }}{\tau }%
\right) \sin \frac{(2k+1)\pi \varphi ^{\prime }}{\varphi _{0}}}rdr.}
\end{equation*}

\subsection{Application to the CVA computation}

We associate the process $x_t$ with the protection seller and the process $%
y_t$ with the reference name issuer of a CDS. The protection buyer will be
considered non-risky in this case. The pricing equation for computing the
CVA is given by:

\begin{equation}
V_{t}+\frac{1}{2}V_{xx}+\frac{1}{2}V_{yy}+\rho _{xy}V_{xy}-\varrho V=0,
\end{equation}%
with the final condition $V(T,x,y)=0$. Boundary conditions are 0, except in
the case when the protection seller defaults first and there will be a
shortfall equal to a fraction of the outstanding present value of the
standard single name swap: 
\begin{equation*}
V^{\textrm{CVA}}(t,0,y)=\left( 1-R_{PS}\right) V^{CDS}\left( t,y\right) ^{+}.
\end{equation*}%
where $R_{PS}$ is the recovery of the protection seller. In order to solve
this problem we apply similar changes of function and variables as in
section \ref{sect:2D_PrEq} and obtain a similar pricing equation 
\begin{equation}
U_{t}+\frac{1}{2}\left( U_{rr}+\frac{1}{r}U_{r}+\frac{1}{r^{2}}U_{\varphi
\varphi }\right) =0,  \label{eq:finally_to_solve}
\end{equation}%
with the final condition: $U(T,r,\varphi )=0$ and boundary conditions: 
\begin{align*}
& U(t,0,\varphi )=0,\quad U(t,\infty ,\varphi )=0,\quad U\left( t,r,0\right)
=0, \\
& U^{\textrm{CVA}}(t,r,\varphi _{0})=e^{\varrho (T-t)}\left( 1-R_{PS}\right) V^{CDS}\left(
t,\bar{\rho}_{xy}r\right) ^{+}.
\end{align*}
The solution for this problem is given by: 
\begin{align*}
U(t,r^{\prime },\varphi ^{\prime })=& \frac{1}{2}\left[ \int_{t}^{T}{%
\!\!\!\int_{0}^{\infty }{\left. \frac{\partial G(t^{\prime }-t,r,\varphi )}{%
\partial \varphi }\right\vert _{\varphi =0}U(t^{\prime },r,0)\frac{1}{r}dr}%
dt^{\prime }}\right. \\
& \left. -\int_{t}^{T}{\!\!\!\int_{0}^{\infty }{\left. \frac{\partial
G(t^{\prime }-t,r,\varphi )}{\partial \varphi }\right\vert _{\varphi
=\varphi _{0}}U(t^{\prime },r,\varphi _{0})\frac{1}{r}dr}dt^{\prime }}\right]
.
\end{align*}
Supplying the boundary conditions for the CVA problem we obtain: 
\begin{equation}
V^{\textrm{CVA}}(t,r^{\prime },\varphi ^{\prime })\!=\!-\textstyle{\frac{1-R_{PS}}{2}%
\!\int\limits_{t}^{T}{\!\int\limits_{0}^{\infty }{\!\left. \frac{\partial
G(t^{\prime }-t,r,\varphi )}{\partial \varphi }\right\vert _{\varphi
=\varphi _{0}}\!\!\!e^{-\varrho (t^{\prime }-t)}V^{CDS}\!\!\left( t^{\prime
},\bar{\rho}_{xy}r\right) ^{+}\frac{1}{r}dr}dt^{\prime }.}}
\end{equation}
Thus, by using Green's function, we can perform the CVA computation in a natural and straightforward way.

\section{Three dimensional case}

\label{sect:3D}

For the three dimensional problem we need to model the dynamics of the asset
values of the reference name, protection seller and protection buyer
simultaneously. As shown previously, we consider the default barrier to be a
deterministic function of time for all three assets and we work directly
with the processes $x_{t}$, $y_{t}$ and $z_{t}$ which measure the
\textquotedblleft relative\textquotedblright\ distance from the default
barrier in time for each of the three entities considered. These processes
have the following dynamics: $dx_{t}=dW_{t}^{x}$, $dy_{t}=dW_{t}^{y}$, $%
dz_{t}=dW_{t}^{z}$, where we correlate the Brownian motions with
correlations $\rho _{xy}$, $\rho _{xz}$, $\rho _{yz}$.

\subsection{Pricing problem}

\label{sect:PricingEq}

The general pricing problem in the $\mathbb{R}_{+}^{3}$ octant is: 
\begin{equation*}
V_{t}+\frac{1}{2}V_{xx}+\frac{1}{2}V_{yy}+\frac{1}{2}V_{zz}+\rho
_{xy}V_{xy}+\rho _{xz}V_{xz}+\rho _{yz}V_{yz}-\varrho V=0.
\end{equation*}%
We consider the following function $U(t,x,y,z)=e^{\varrho (T-t)}V(t,x,y)$,
and introduce a change of variables that allows us to eliminate the cross
derivatives: 
\begin{equation}
\alpha =x,\ \ \ \beta =\frac{\left( -\rho _{xy}x+y\right) }{\bar{\rho}_{xy}}%
,\ \ \ \gamma =\frac{\left( \left( \rho _{xy}\rho _{yz}-\rho _{xz}\right)
x+\left( \rho _{xy}\rho _{xz}-\rho _{yz}\right) y+\bar{\rho}_{xy}z\right) }{%
\bar{\rho}_{xy}\chi },  \label{eq:changeOfVars1}
\end{equation}%
where we use the notation: $\chi =\sqrt{1-\rho _{xy}^{2}-\rho
_{xz}^{2}-\rho _{yz}^{2}+2\rho _{xy}\rho _{xz}\rho _{yz}}$.

With the change of variables, we have also changed the domain in which we
need to solve the pricing problem. The domain becomes the volume bounded by the planes: $%
\alpha =0$, $\left( \alpha ,-\frac{\rho _{xy}}{\overline{\rho }_{xy}}\alpha
,\gamma \right) $ and $\left( \alpha ,\beta ,\frac{\overline{\rho }_{xy}}{%
\chi }\left( -\rho _{xz}\alpha +\frac{\rho _{xy}\rho _{xz}-\rho _{yz}}{%
\overline{\rho }_{xy}}\beta \right) \right) $. In order to take advantage of
the symmetry of the problem we perform a second change of variables to
spherical coordinates: the axis $\alpha =0$ and $\beta =0$ is given by $%
\theta =0$; the axis $\alpha =0$ and $\gamma =0$ is given by $\varphi =0$
and $\theta =\pi /2$, so that $\left( \alpha ,\beta ,\gamma \right) =\left(
r\sin \theta \sin \varphi ,r\sin \theta \cos \varphi ,r\cos \theta \right) .$

\begin{figure}[b]
\begin{minipage}{0.4\textwidth}
\caption{Domain after the change in coordinates for $\rho_{xy} = 20\%$, $\rho_{xz} = 0\%$, $\rho_{yz} = 30\%$\label{fig:3D_Domain}}
\end{minipage}
\hfill 
\begin{minipage}{0.55\textwidth}
	\centering
	\includegraphics[width = 0.6\textwidth]{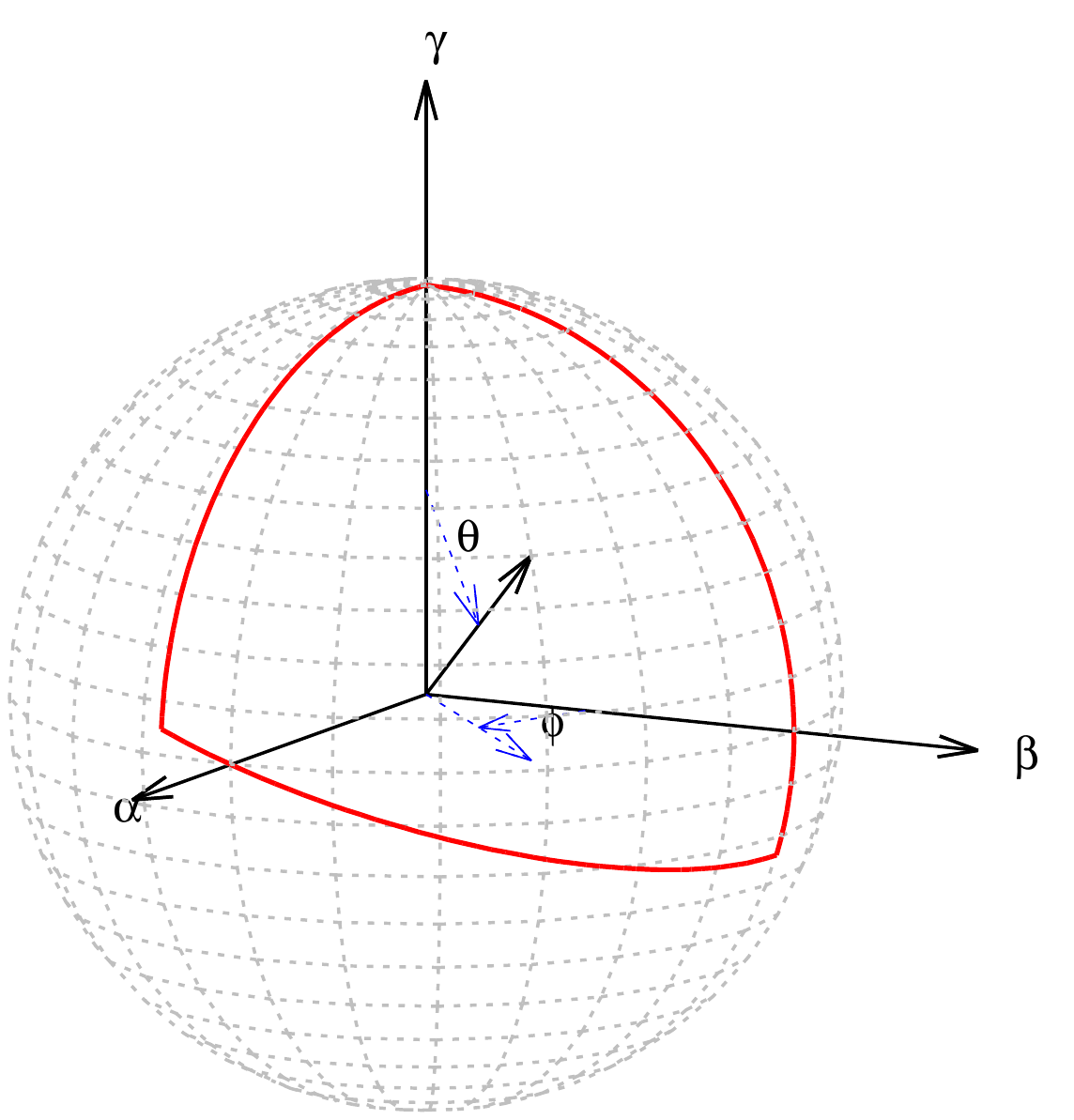}
\end{minipage}
\end{figure}

The range of values for $\varphi $ is given by: $0\leq \varphi \leq \varphi
_{0}$ where \linebreak $\varphi _{0}~=~\arccos \left( -\rho _{xy}\right) $.
As can be observed in figure \ref{fig:3D_Domain}, the possible range of
values for $\theta $ will depend on $\varphi $, so we have $0\leq \theta
\leq \Theta (\varphi )$. Formulas \eqref{eq:ParamPhi} and %
\eqref{eq:ParamTheta} give a parametric characterization of this boundary of
the domain which will prove very useful going forward.

\begin{equation}
\varphi \left( \omega \right) =\arccos \left( \frac{1-\rho _{xy}\omega }{%
\sqrt{1-2\rho _{xy}\omega +\omega ^{2}}}\right) ,  \label{eq:ParamPhi}
\end{equation}%
\begin{equation}
\Theta \left( \omega \right) =\arccos \left( -\frac{\rho _{yz}-\rho
_{xz}\rho _{xy}+\omega \left( \rho _{xz}-\rho _{yz}\rho _{xy}\right) }{\sqrt{%
\overline{\rho }_{xy}\left( \overline{\rho }_{xz}^{2}-2\omega \left( \rho
_{xy}-\rho _{xz}\rho _{yz}\right) +\omega ^{2}\overline{\rho }%
_{yz}^{2}\right) }}\right) .  \label{eq:ParamTheta}
\end{equation}%
In the domain described above the final form of the pricing equation is: 
\begin{equation}
U_{t}+\frac{1}{2}\left[ \frac{1}{r}\frac{\partial ^{2}}{\partial r^{2}}%
\left( rU\right) +\frac{1}{r^{2}}\left( \frac{1}{\sin ^{2}\theta }U_{\varphi
\varphi }+\frac{1}{\sin \theta }\frac{\partial }{\partial \theta }\left(
\sin \theta U_{\theta }\right) \right) \right] =0,  \label{eq:3D_PrEq}
\end{equation}%
with appropriate boundary conditions depending on the payoff.

\subsection{Green's function}

\label{sect:3D_Greens}

We now concentrate on solving the forward problem for the Green's function
in spherical coordinates: 
\begin{eqnarray}
&&G_{\tau }-\frac{1}{2}\left[ \frac{1}{r}\frac{\partial ^{2}}{\partial r^{2}}%
\left( rG\right) +\frac{1}{r^{2}}\left( \frac{1}{\sin ^{2}\theta }G_{\varphi
\varphi }+\frac{1}{\sin \theta }\frac{\partial }{\partial \theta }\left(
\sin \theta G_{\theta }\right) \right) \right] =0,  \label{eq:3D_Greens} \\
&&G(0,r,\varphi ,\theta )=\frac{1}{r^{2}\sin \theta }\delta \left(
r-r^{\prime }\right) \delta \left( \varphi -\varphi ^{\prime }\right) \delta
\left( \theta -\theta ^{\prime }\right) ,  \notag \\
&&G\left( \tau ,r,0,\theta \right) =G(\tau ,r,\varphi _{0},\theta )=G\left(
\tau ,r,\varphi ,0\right) =0,  \notag \\
&&G(\tau ,r,\varphi ,\Theta (\varphi ))=G(\tau ,0,\varphi ,\theta )=0,\quad
G(\tau ,r,\varphi ,\theta )\xrightarrow[r\to \infty]{}0.  \notag
\end{eqnarray}

We aim to build a solution for the Green's function through the eigenvalues
expansion method. The first step is to apply the separation of variables
technique: 
\begin{equation}
G(\tau ,r,\varphi ,\theta )=g(\tau ,r)\Psi (\varphi ,\theta ).
\label{eq:G_separation}
\end{equation}%
By substituting \eqref{eq:G_separation} in \eqref{eq:3D_Greens} we obtain an equation where the left hand side depends only on $\tau$ and $r$ and the right hand side depends only on $\varphi$ and $\theta$ and hence both sides are equal to some constant value $C$, which is necessarily negative; we use the notation $C = -\Lambda^2$.

For function $g(\tau ,r)$ we have the initial condition $g(0,r)=\frac{1}{%
r^{2}}\delta \left( r-r^{\prime }\right) ,$ and boundary conditions $g(\tau
,0)=0$ and $g(\tau ,r)\xrightarrow[r\to \infty]{}0$. Function $g$ solves the following PDE 
\begin{equation*}
g_{\tau }=\frac{1}{2}\left( \frac{1}{r}\frac{\partial ^{2}}{\partial r^{2}}%
\left( rg\right) -\frac{\Lambda ^{2}}{r^{2}}g\right) ,
\end{equation*}%
which is similar to the equation utilised for the two dimensional
case. The solution is given by: 
\begin{equation*}
g(\tau ,r)=\frac{e^{-\frac{r^{2}+r^{\prime 2}}{2\tau }}}{\tau \sqrt{%
rr^{\prime }}}I_{\sqrt{\Lambda ^{2}+1/4}}\left( \frac{rr^{\prime }}{\tau }%
\right) .
\end{equation*}

In order to obtain the Green's function we also need to solve the angular
part PDE:%
\begin{equation}
\frac{1}{\sin ^{2}\theta }\Psi _{\varphi \varphi }+\frac{1}{\sin \theta }%
\frac{\partial }{\partial \theta }\left( \sin \theta \Psi _{\theta }\right)
=-\Lambda ^{2}\Psi ,  \label{eq:sphericalLaplacian}
\end{equation}%
with zero boundary conditions: $\Psi (0,\theta )=\Psi (\varphi _{0},\theta
)=\Psi (\varphi ,0)=\Psi (\varphi ,\Theta (\varphi ))=0$. It is well-known
that the spectrum of this problem is discrete and the set of the
corresponding eigenvectors is complete.

The two dimensional spherical surface inside the red line shown in figure %
\ref{fig:3D_Domain} may be mapped directly onto the $\left(\varphi,\theta%
\right)$ plane. This is done in a similar way to the method used by cartographers to map the Earth's
surface using Mercator's projection. The southern boundary of the domain is
mapped into a continuous curve parametrised by equations \eqref{eq:ParamPhi}
and \eqref{eq:ParamTheta}. The boundary at $\theta = 0$ is degenerate as it
corresponds to the north pole on the sphere. Figure \ref{fig:2D_30_50_80}
and \ref{fig:2D__45__65_80} show the domain (denoted by $\Omega$) projected onto the $%
\left(\varphi,\theta\right)$ plan for sample correlation values.

\noindent \vspace{-0.2cm} 
\begin{minipage}{0.45\textwidth}
\domain{0.3}{0.5}{0.8}{2D_30_50_80}
\end{minipage}
\hfill 
\begin{minipage}{0.45\textwidth}
\domain{-0.45}{-0.65}{0.8}{2D__45__65_80}
\end{minipage}
\vspace{1cm}

In our case we are interested in writing an eigenvalue expansion for the
Green's function.The weak formulation for our problem is given by 
\begin{equation}
\int_{\Omega }{\frac{1}{\sin \theta }\Psi _{\varphi }\Psi _{\varphi
}^{\prime }d\Omega }+\int_{\Omega }{\sin \theta \Psi _{\theta }\Psi _{\theta
}^{\prime }d\Omega }=\Lambda ^{2}\int_{\Omega }{\Psi \Psi ^{\prime }\sin
\theta d\Omega },  \label{eq:weakFormulation}
\end{equation}%
where $\Psi ^{\prime }$ is a test function that belongs to the same space as 
$\Psi$, in particular it is also $0$ on the boundary of the domain.

The first step necessary is to construct a mesh on the domain of interest.
We construct a triangular mesh following the ideas presented in %
\citet{MeshThesis}. For the actual mesh generation, the algorithm uses an
iterative technique. We use adaptive grids that are finer along the
boundaries.

\begin{figure}[htbp]
\centering
\makebox[\textwidth]{
  \subfigure[First iteration mesh]
  {
  	\includegraphics[width = 0.5\textwidth]{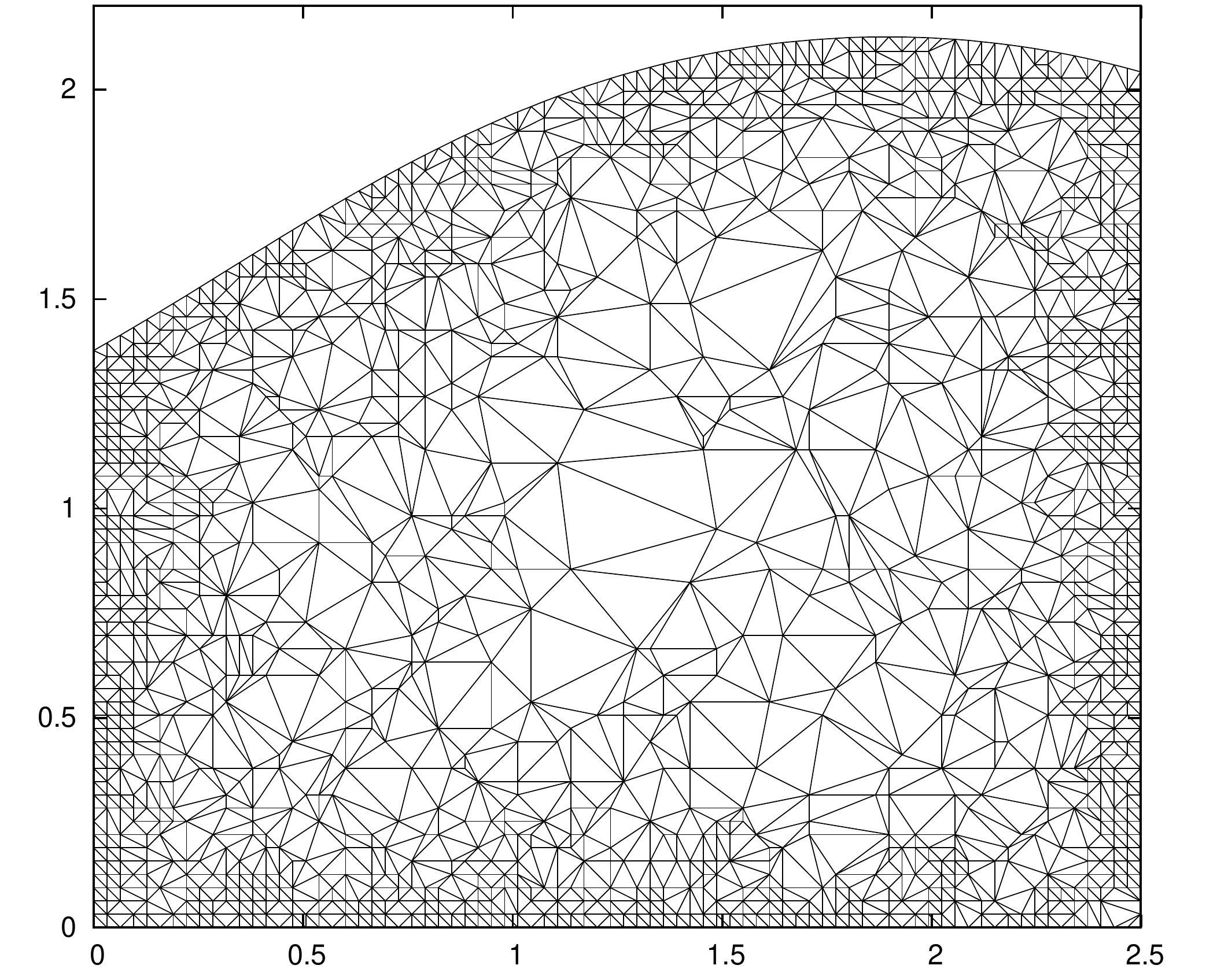}
  }
  \subfigure[Mesh after 100 iterations]
  {
  	\includegraphics[width = 0.5\textwidth]{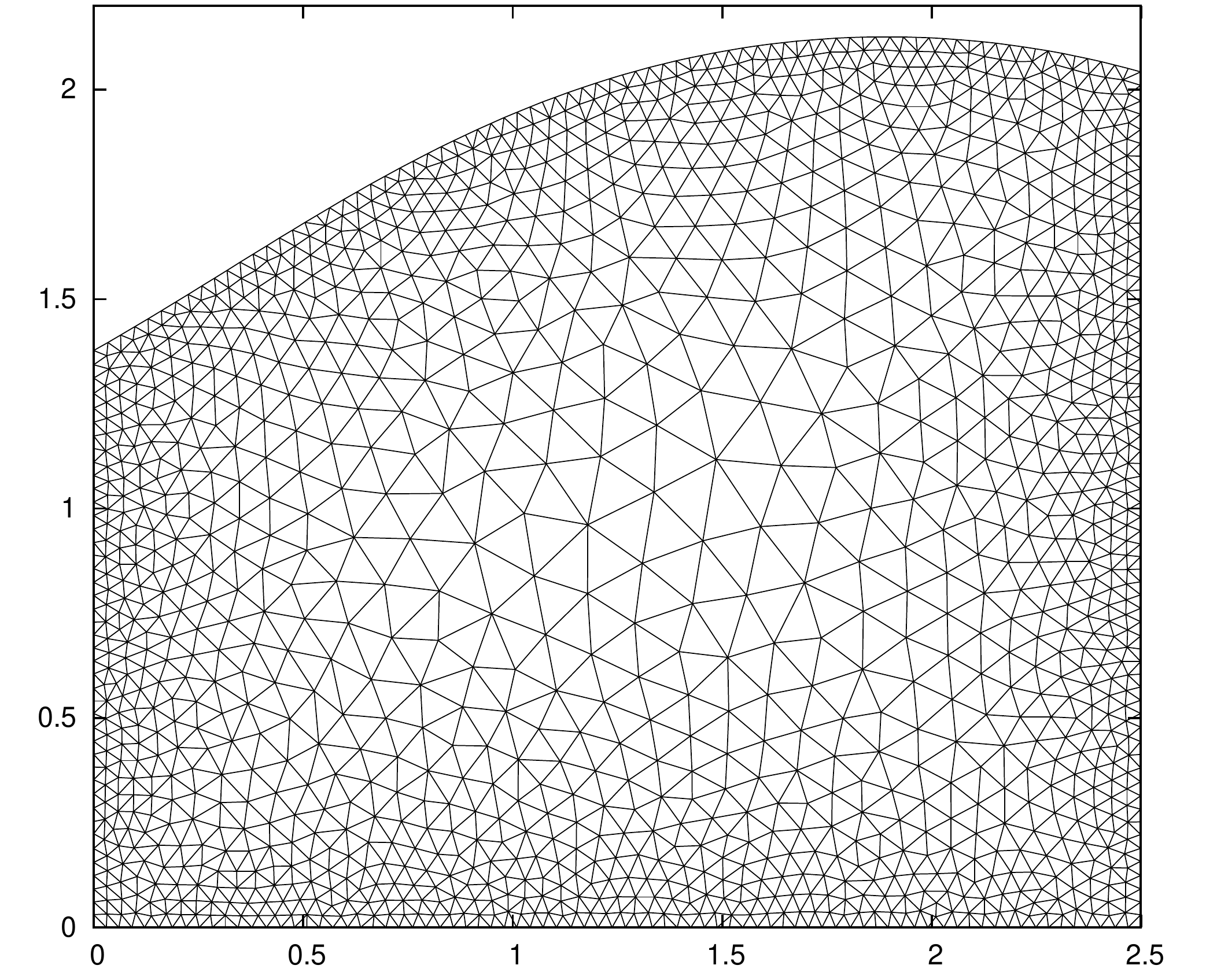}
  }}
\caption{Adaptive mesh for the domain obtained for $\protect\rho_{xy} = 80\%$%
, $\protect\rho_{xz} = 50\%$, $\protect\rho_{yz} = 30\%$. The mesh is
constructed using 1500 points and is finer as we get closer to the
boundaries.}
\label{fig:Meshes_AllB_80_50_30}
\end{figure}

Figure \ref{fig:Meshes_AllB_80_50_30} shows an example of mesh obtained for
sample values of the correlations. We show the initial mesh and the final mesh obtained after 100 iterations. 

Once the mesh is constructed we solve the eigenvalue problem in matrix form
and obtain the eigenvalues and corresponding eigenvectors. Figure \ref%
{fig:EV_80_20_50} shows sample eigenvectors for a domain where all three
correlations are positive. 
\begin{figure}[htbp]
\centering
\makebox[\textwidth]{
  \subfigure[Eigenvector 1: $\Lambda_{1}^2 = 5.2$]
  {
  	\includegraphics[width = 0.5\textwidth]{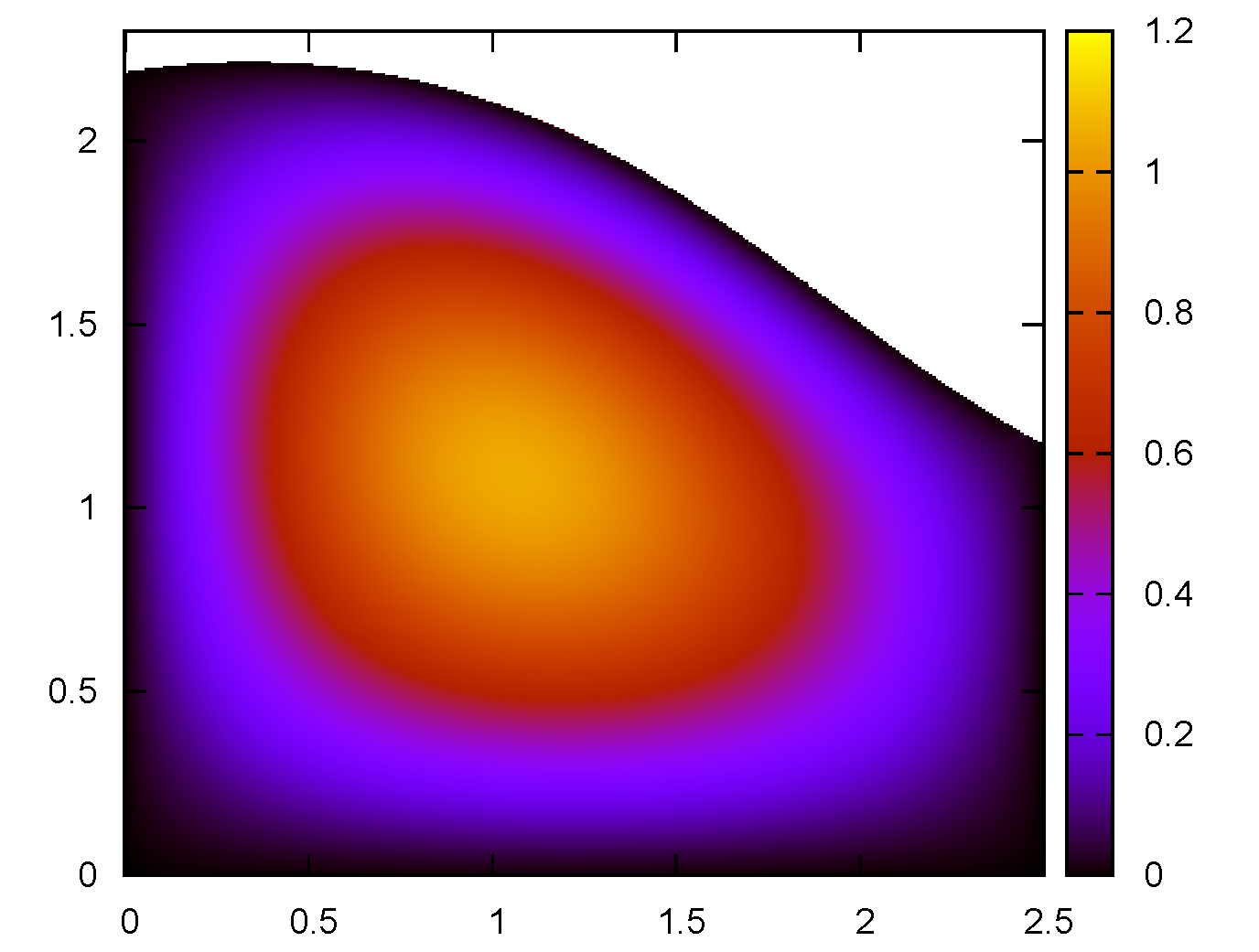}
  }
  \subfigure[Eigenvector 3: $\Lambda_{3}^2 = 16.3$]
  {
  	\includegraphics[width = 0.5\textwidth]{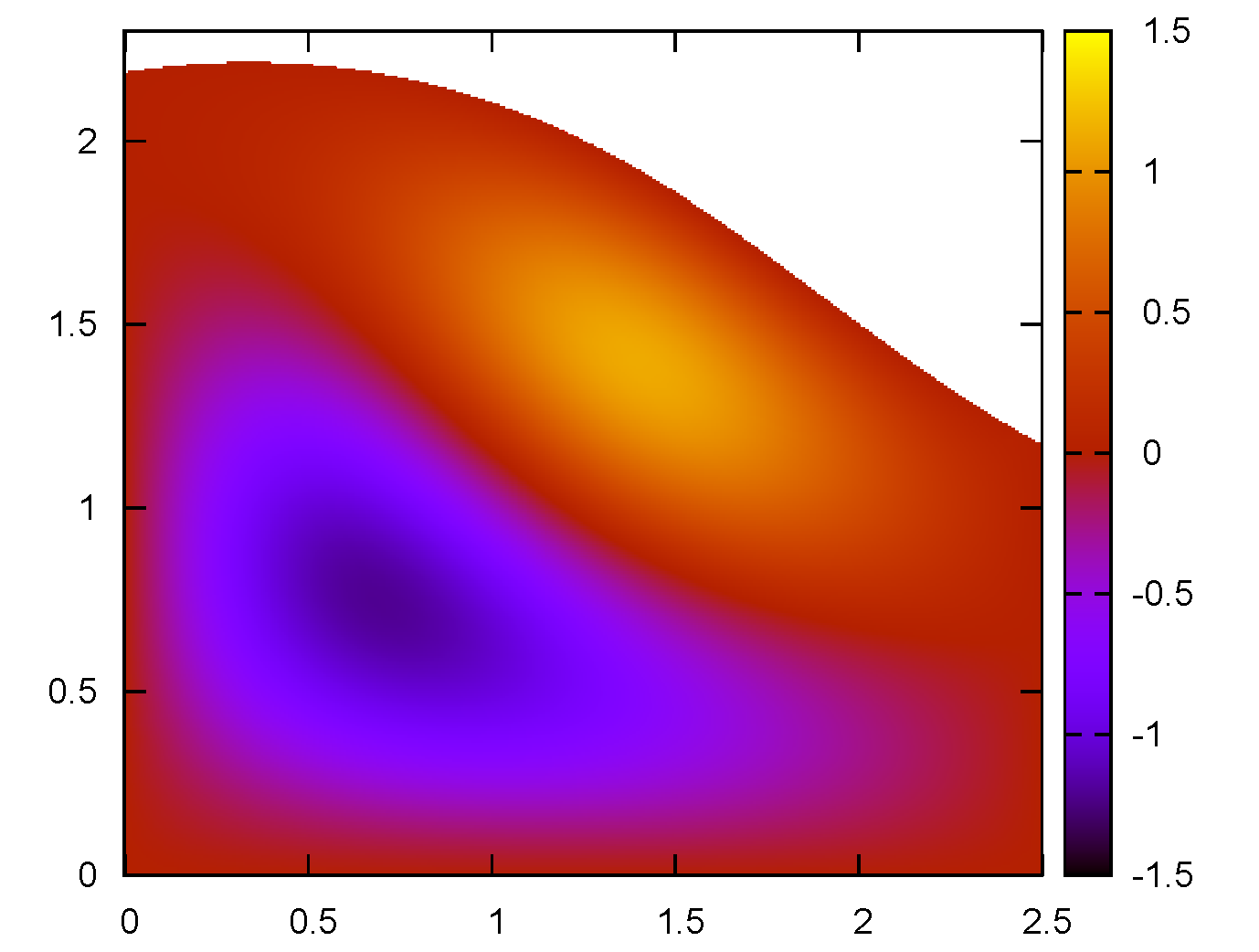}
  }} 
\makebox[\textwidth]{
  \subfigure[Eigenvector 4: $\Lambda_{4}^2 = 21.3$]
  {
  	\includegraphics[width = 0.5\textwidth]{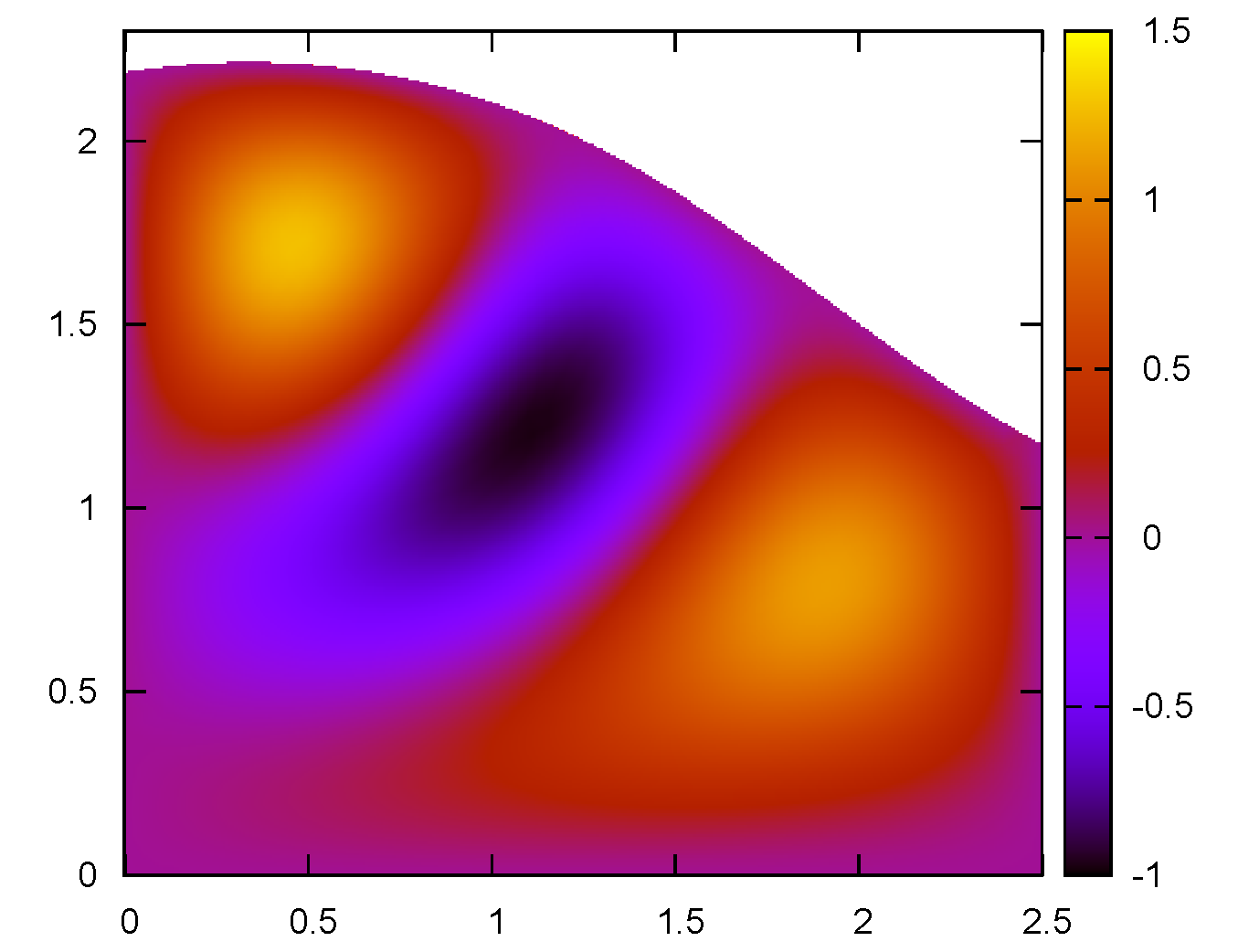}
  }  \subfigure[Eigenvector 30: $\Lambda_{30}^2 = 140.0$]
  {
  	\includegraphics[width = 0.5\textwidth]{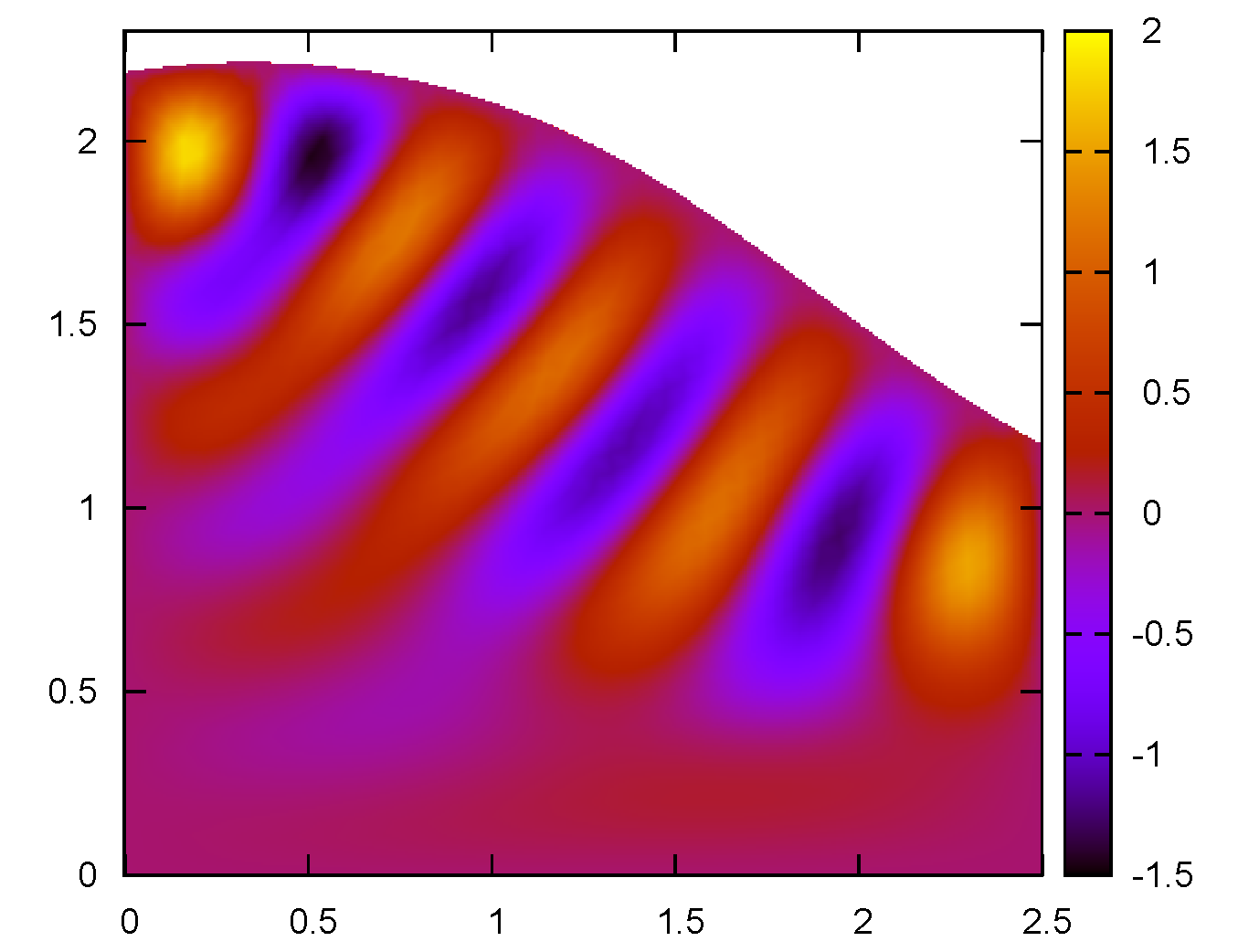}
  }}
\caption{Eigenvectors and corresponding eigenvalues for the domain obtained
for $\protect\rho_{xy} = 80\%$, $\protect\rho_{xz} = 20\%$, $\protect\rho%
_{yz} = 50\%$.}
\label{fig:EV_80_20_50}
\end{figure}

Having calculated the eigenvectors and eigenvalues for our problem we can
write the eigenfunction expansion for $\Psi \left( \varphi ,\theta \right)$,
and then for the Green's function we obtain the following final formula: 
\begin{equation}
G\left( \left. \tau ,r,\varphi ,\theta \right\vert r^{\prime },\varphi
^{\prime },\theta ^{\prime }\right) =\frac{e^{-\frac{r^{2}+r^{\prime 2}}{%
2\tau }}}{\tau \sqrt{rr^{\prime }}}\sum_{n=1}^{\infty }{I_{\sqrt{\Lambda
_{n}^{2}+\frac{1}{4}}}\left( \frac{rr^{\prime }}{\tau }\right) \Psi
_{n}(\varphi ^{\prime },\theta ^{\prime })\Psi _{n}(\varphi ,\theta ).}
\label{eq:Greens}
\end{equation}

\subsection{Joint survival probability}

Similarly to the two dimensional case, we denote by $Q(t,x,y,z)$ the joint
survival probability of issuers $x$, $y$ and $z$ to a fixed maturity $T$.
This solves the following pricing equation 
\begin{equation}
Q_{t}+\frac{1}{2}Q_{xx}+\frac{1}{2}Q_{yy}+\frac{1}{2}Q_{zz}+\rho
_{xy}Q_{xy}+\rho _{xz}Q_{xz}+\rho _{yz}Q_{yz}=0
\end{equation}%
with final condition $Q(T,x,y,z)=1$ and zero boundary conditions. We proceed
with a similar change of variables as described in section \ref{sect:PricingEq},
and using the expression for the Green's function given in equation %
\eqref{eq:Greens} we obtain: 
\begin{equation*}
Q(t,r^{\prime },\varphi ^{\prime },\theta ^{\prime
})=\int\limits_{0}^{\infty }\ {\frac{e^{-\frac{r^{2}+r^{\prime 2}}{2\tau }}}{%
{\tau }\sqrt{r^{\prime }}}\sum_{n=1}^{\infty }{\!I_{\scriptscriptstyle\sqrt{%
\Lambda _{n}^{2}+\frac{1}{4}}}\!\left( \frac{rr^{\prime }}{{\tau }}\right) \!%
{\scriptstyle{\Psi _{n}(\varphi ^{\prime },\theta ^{\prime })}}\!\left[ \int 
{\!\!\!\!\int_{\Omega }{\!\!\scriptstyle{\Psi _{n}(\varphi ,\theta )\sin
\theta d\Omega }}}\right] }r^{\frac{3}{2}}dr.}
\end{equation*}

\subsection{Application to the CVA computation}

We associate the process $x_{t}$ with the protection seller, the process $%
y_{t}$ with the reference name and $z_{t}$ with the protection buyer. The
pricing equation for computing CVA or DVA in the case where all three names
are risky is given by: 
\begin{equation}
V_{t}+\frac{1}{2}V_{xx}+\frac{1}{2}V_{yy}+\frac{1}{2}V_{zz}+\rho
_{xy}V_{xy}+\rho _{xz}V_{xz}+\rho _{yz}V_{yz}-\varrho V=0,
\end{equation}%
with the final condition $V(T,x,y,z)=0$ and boundary conditions depending on
the payoff.

In the case of the CVA calculation, we get a payout if the protection seller
defaults, 
the payout is: 
\begin{equation}
V^{\text{CVA}}(t,0,y,z)=\left( 1-R_{PS}\right) V(t,y)^{+},
\end{equation}%
where $V(t,y)^{+}$ is the positive value of the single name default swap
with non-risky counterparts at the time of the default of the protection
seller.

Similarly we have the payout for the DVA calculation: 
\begin{equation}
V^{\text{DVA}}(t,x,y,0)=\left( 1-R_{PB}\right) V(t,y)^{-},
\end{equation}%
where $R_{PB}$ is the recovery of the protection buyer and $V(t,y)^{-}$ is
the negative value of the single name default swap with non-risky
counterparts at the time of the default of the protection buyer. For both
CVA and DVA calculations the boundary conditions are 0 for all other cases.
Following the same procedure as in section \ref{sect:PricingEq} we obtain
the modified pricing equation 
\begin{equation}
U_{t}+\frac{1}{2}\left[ \frac{1}{r}\frac{\partial ^{2}}{\partial r^{2}}%
\left( rU\right) +\frac{1}{r^{2}}\left( \frac{1}{\sin ^{2}\theta }U_{\varphi
\varphi }+\frac{1}{\sin \theta }\frac{\partial }{\partial \theta }\left(
\sin \theta U_{\theta }\right) \right) \right] =0,  \label{eq:pricingEq}
\end{equation}%
with final condition $U(T,r,\varphi ,\theta )=0$ and 0 boundary conditions
except for: 
\begin{align}
U^{\scriptscriptstyle{\text{CVA}}}\!(t,r,0,\theta )\!=& e^{\varrho (T-t)}\!\left( 1-R_{\scriptscriptstyle{PS}}\right) V(t,%
\overline{\rho }_{xy}r\sin \theta )^{+},  \label{eq:CVABoundaryCond} \\
U^{\scriptscriptstyle{\text{DVA}}}\!\left( t,r,\varphi ,\theta \left( \varphi \right) \right)\! =&
e^{\varrho (T-t)}\!\left( 1-R_{\scriptscriptstyle{PB}}\right) V\left( t,\textstyle{%
r\sin \theta \left( \rho _{xy}\sin \varphi +\overline{\rho }_{xy}\cos
\varphi \right) }\right)^{-}  \label{eq:DVABoundaryCond}
\end{align}%
for the CVA and DVA respectively.

The final pricing formula for $U$ is given by: 
\begin{align}
U\left( t,r^{\prime}\!\!,\varphi^{\prime}\!\!,\theta ^{\prime }\right) =& -\frac{1%
}{2}\int\limits_{t}^{T}{\!\!\int\limits_{0}^{\infty }{\!\int\limits_{0}^{%
\varphi _{0}}{\!\!\sin \theta \left( \varphi \right) \,U\left( t^{\prime
},r,\varphi ,\theta \left( \varphi \right) \right) \,G_{\theta }\left(
t^{\prime },r,\varphi ,\theta \left( \varphi \right) \right) \,d\varphi }dr}%
dt^{\prime }}  \notag \\
& +\frac{1}{2}\!\!\int\limits_{t}^{T}{\!\!\!\int\limits_{0}^{\infty }{\!\!\int\limits_{0}^{\infty
}{\!\!\frac{U \left( t^{\prime}\!,r,\varphi \left(\omega\right)\!,\theta \left( \omega \right) \right)G_{\varphi }\left( t^{\prime}\!,r,\varphi \left(\omega\right)\!,\theta \left( \omega \right) \right)}{\sin \theta \left( \omega \right) } \theta^{\prime}\left( \omega \right)
d\omega }dr}dt^{\prime }}  \notag \\
& -\frac{1}{2}\int_{t}^{T}{\!\!\!\int_{0}^{\infty }{\!\!\!\int_{0}^{\theta
\left( \varphi _{0}\right) }{\!\!\frac{U\left( t^{\prime
},r,\varphi _{0},\theta \right) \,G_{\varphi }\left( t^{\prime },r,\varphi
_{0},\theta \right)}{\sin \theta } d\theta }dr}dt^{\prime }}  \notag \\
& +\frac{1}{2}\int_{t}^{T}{\!\!\!\int_{0}^{\infty }{\!\!\!\int_{0}^{\theta
(0)}{\!\!\frac{U\left( t^{\prime },r,0,\theta \right)
\,G_{\varphi }\left( t^{\prime },r,0,\theta \right)}{\sin \theta } d\theta }dr}dt^{\prime }.%
}  \label{eq:pricingFormula}
\end{align}

We note that for one of the integrals above we have used the parametric
representation of the boundary of our domain given by formulas %
\eqref{eq:ParamPhi} and \eqref{eq:ParamTheta} To obtain the precise formulas
for the CVA and DVA calculations we now apply the boundary conditions in
equations \eqref{eq:CVABoundaryCond} and \eqref{eq:DVABoundaryCond}
respectively: 
\begin{equation}
U^{\scriptscriptstyle{\textrm{CVA}}}\!\left( t,r^{\prime }\!\!,\varphi ^{\prime }\!\!,\theta ^{\prime
}\right) =\frac{1}{2}\int\limits_{t}^{T}{\!\!\int\limits_{0}^{\infty }{%
\!\int\limits_{0}^{\theta (0)}{\!\!\frac{U^{\scriptscriptstyle{\textrm{CVA}}%
}(t^{\prime },r,0,\theta )G_{\varphi }\left( t^{\prime },r,0,\theta \right)}{\sin \theta }d\theta }dr}dt^{\prime
},}
\end{equation}%
\begin{align}
U^{\scriptscriptstyle{\textrm{DVA}}}\!\left( t,r^{\prime}\!\!,\varphi^{\prime}\!\!,\theta ^{\prime
}\right)\! =& -\frac{1}{2}\!\int\limits_{t}^{T}{\!\!\int\limits_{0}^{\infty }{%
\!\int\limits_{0}^{\varphi _{0}}{\!\!\sin\! \theta\!\left(\varphi\right)
U^{\scriptscriptstyle{\textrm{DVA}}}\!\!\left( t^{\prime },r,\varphi
,\theta \left( \varphi \right) \right)G_{\theta}\left( t^{\prime },r,\varphi
,\theta \left( \varphi \right) \right) \!d\varphi }dr}dt^{\prime }} \nonumber \\
& +\!\frac{1}{2}\!\!\int\limits_{t}^{T}{\!\!\int\limits_{0}^{\infty }{%
\!\!\int\limits_{0}^{\infty }{\!\!\frac{U^{\scriptscriptstyle{\text{DVA}}}\!\!\left( t^{\prime}\!,r,\varphi \!\left( \omega \right) \!,\theta \!\left( \omega \right)
\right)G_{\varphi }\!\!\left( t^{\prime}\!,r,\varphi \!\left( \omega \right) \!,\theta \!\left( \omega \right)
\right)}{\sin \theta \left(
\omega \right) } \theta ^{\prime }\!\left( \omega \right)\! d\omega }dr}dt^{\prime }.}
\end{align}
These original formulas provide a new way of consistently computing the CVA and DVA. Similar ideas can be used for many other purposes, which will be discussed elsewhere.


\section{Numerical results}

\label{sect:results}

In this section we present the results of the CVA and DVA calculations for a
risky CDS. We compare the breakeven coupon obtained for a standard CDS to
those obtained when either the protection buyer or the protection seller
are risky (using the 2D formulation and results), as well as when both are
risky (using the 3D formulation and results). When using the 2D formulation
and considering that either the protection seller or the protection buyer
are risky, the two parties will not agree on the breakeven coupon of the
CDS. This problem goes away when using the full three dimensional framework
in which both are risky and the problem becomes symmetrical.

We consider three issuers for our example: AIG as a protection seller, GE as
a reference name of the CDS, and UNICREDIT a protection buyer. We have
chosen sufficiently risky entities for the protection seller and the
protection buyer$\ $in order to emphasize the effect of the CVA and DVA
adjustments on the break even coupon. We calibrate the model inputs to the
market data from the 15th of December 2011 (see table \ref{tab:inputs}).

\begin{table}[hbp]
\centering
\noindent%
\makebox[\textwidth]{
    \begin{tabular}{ |c||c|c|c|}
        \hline
        Inputs & AIG & GE & UNICREDIT \\
        \hline\hline
        Initial value & 0.0359 & 0.3035 & 0.1199  \\
        $\sigma$ & 2.44\% & 10.45\% & 6.3\% \\
        Recovery & 50\% & 40\% & 40\% \\
        \hline
		\end{tabular}
		}
\caption{Input parameters calibrated to market data (15th December 2011)}
\label{tab:inputs}
\end{table}

The initial value is a measure of the relative distance to default. The
volatility $\sigma$ has been calibrated such that the 5Y single name CDS
spread is matched to the market spread (the 5Y point has been chosen at it
is usually the most liquidly traded contract).

For the 2D and 3D cases we also need the correlations between the different
issuers as inputs to our model. These can be calibrated from the prices of
the first to default swap contracts if such contracts including the relevant
names are available on the market. Alternatively we can proxy these
correlations by assigning a sector to each issuer and then using the sector
to sector historically estimated correlations. In this section however we
aim to show the impact of CVA and DVA on the breakeven spread of a CDS, and
hence we will use different sets of pairwise correlations for the same group
of issuers in order to illustrate a variety of cases.

\begin{figure}[h!]
\begin{minipage}{0.48\textwidth}
  \centering
  	\includegraphics[width = \textwidth]{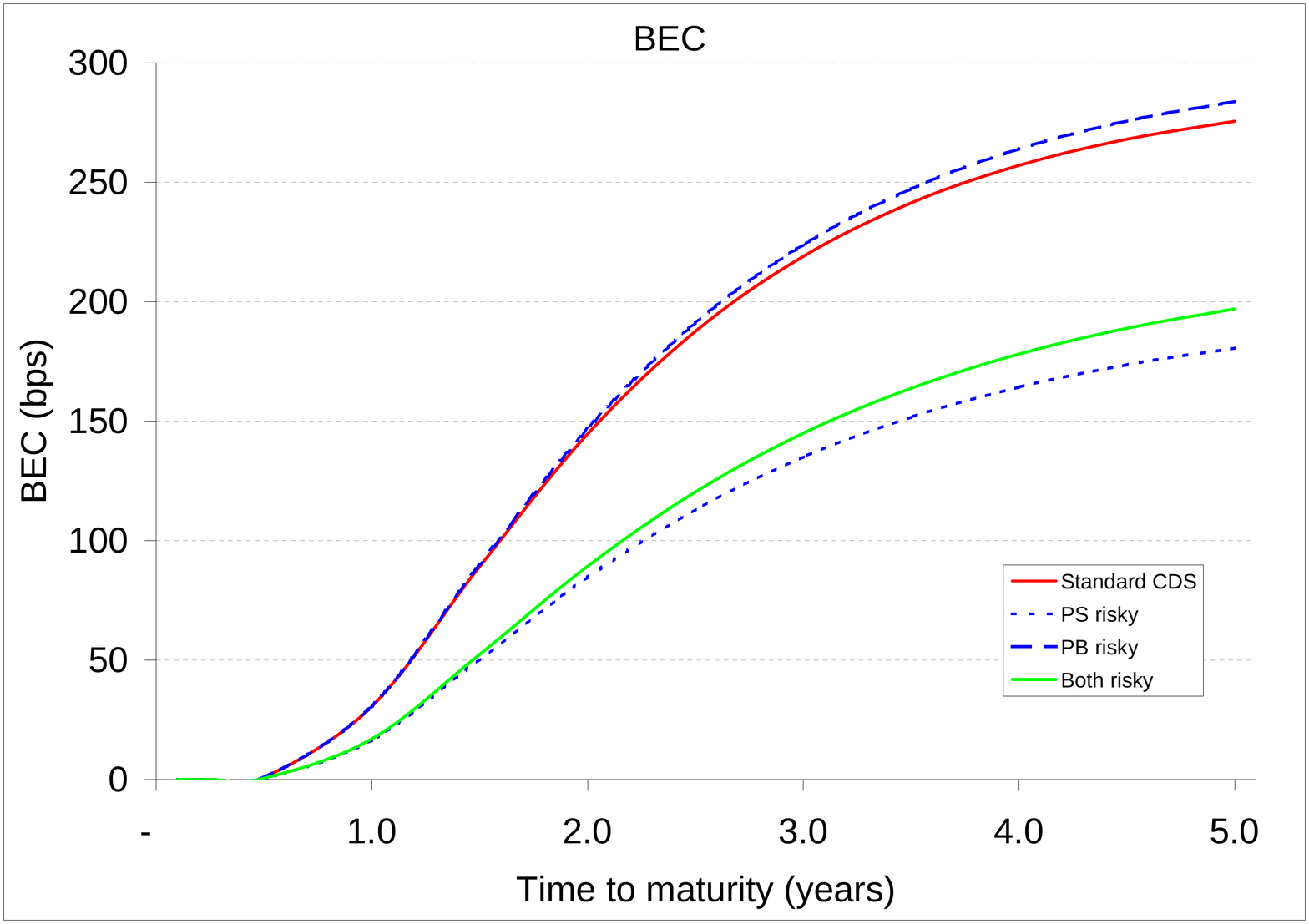}
	\caption{$\rho_{xy} = 80\%$, $\rho_{xz} = 50\%$, $\rho_{yz} = 30\%$}
  \label{fig:Results_80_50_30}
\end{minipage}
\hfill 
\begin{minipage}{0.48\textwidth}
  \centering
  	\includegraphics[width = \textwidth]{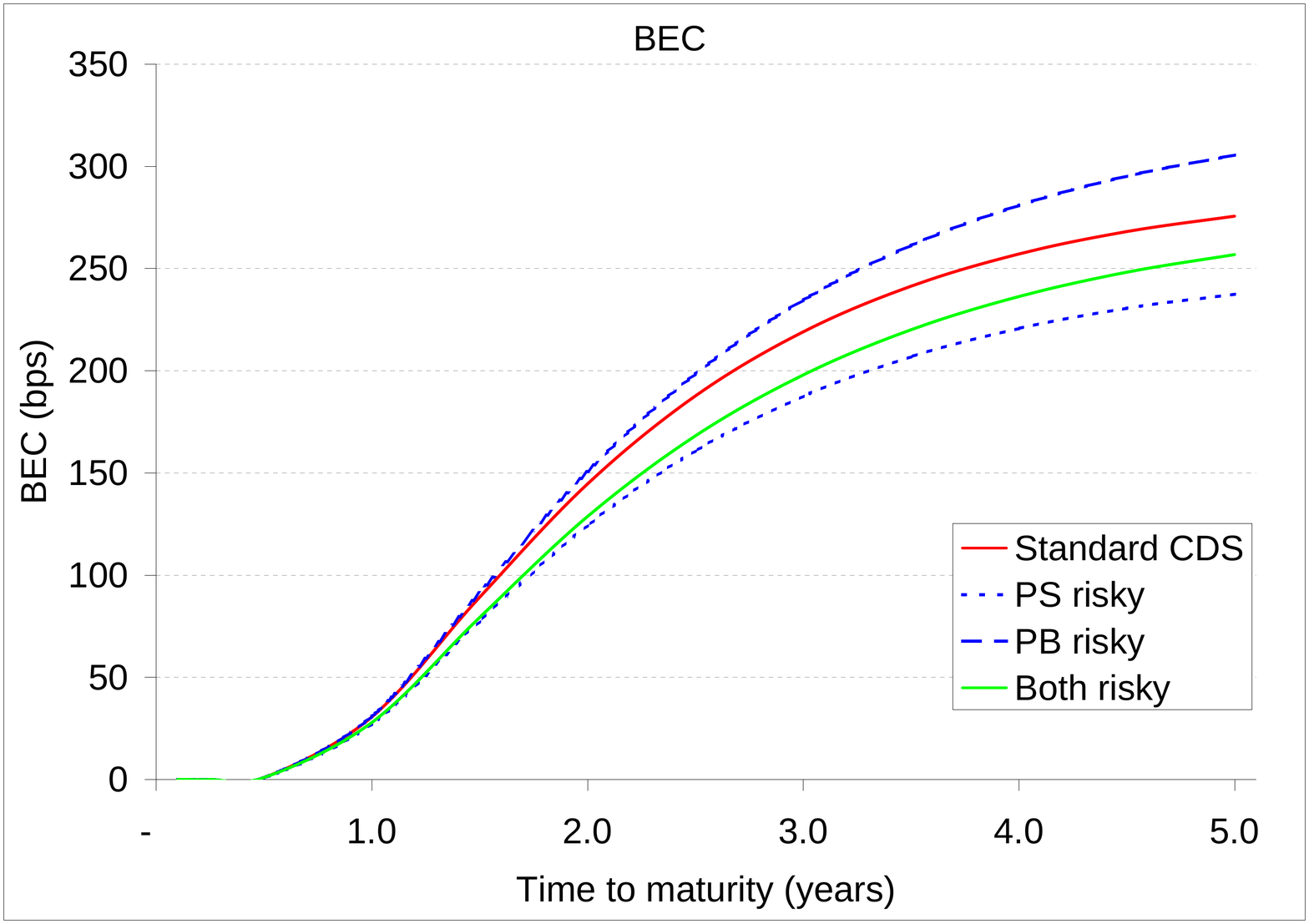}
	\caption{$\rho_{xy} = 20\%$, $\rho_{xz} = -10\%$, $\rho_{yz} = -60\%$}
  \label{fig:Results_20__10__60}
\end{minipage}
\end{figure}

Figure \ref{fig:Results_80_50_30} shows the case where the protection seller
is highly correlated to the reference name. We observe that the spreads are
hyper-exponentially flat at 0, which is a known problem for models without
jumps. However for longer maturities we can match model and market prices.
We use the calibrated model in order to analyse the effects of either the
protection seller, or the protection buyer, or both being risky.

If the protection seller is risky, the probability of it not paying the full
amount due in the case of the default of the reference name is non-zero, and
hence the protection buyer pays a lower coupon as it takes on that risk as
well. If the protection buyer is risky then the breakeven coupon moves in the
opposite direction and the two counterparties no longer agree on the coupon.
Typically, the latter shift is much smaller than the former. In the case
when both are considered risky the problem becomes symmetrical, and a
mutually agreeable coupon can be computed.

In this example, in the case of a default of the reference name, the
protection seller is likely to default as well, and hence the shortfall
between the contractual payout and what will actually be paid can be
significant. The break-even coupon will adjust accordingly and will be
lower than on a standard fully collateralised CDS as the expectation of the
payout is lower from the protection buyer's point of view.

Figure \ref{fig:Results_20__10__60} shows the case where the protection
buyer is highly anti-correlated to the reference name. This is intuitively
the case where the DVA is largest as it is in the cases where the reference
name does not default that the protection buyer is more likely to default on
its coupon paying obligation. This leaves the protection seller with a
potential shortfall.


\section{Conclusion}

\label{sect:conclusion}

A 3D extension of the structural default framework where the joint dynamics
of the firms' values are driven by correlated Brownian motions is presented.
A method to obtain a semi-analytical expression for the Green's function
using the eigenvalue expansion method is developed. It is shown how to apply
this in order to compute joint survival probabilities of three different
companies and how to calculate the credit and debit value adjustments for a
standard CDS. In the 3D case, a fully analytical expression is not
available, since the eigenvalues and eigenvectors have to be computed using
the finite element method. Given a triplet of correlations, however, these can be precomputed, which then allows efficient computations across a range of initial points, volatilities
or other trade-related data (coupons, recoveries etc). Concrete examples demonstrate that the CVA\ and
DVA for a typical CDS can be very large.

\section*{Acknowledgements}
We wish to thank Leif Andersen, Peter Franke, Stewart Inglis, Guillaume Kirsch and Marsha Lipton for helpful discussions and useful comments.

\bibliographystyle{plainnat}
\bibliography{ThesisBib}

\end{document}